\documentclass[useAMS,usenatbib]{mn2e}
\usepackage{natbib}
\usepackage{graphicx}
\usepackage{amssymb}
\usepackage{caption}

\begin{document}
\date{Accepted 2014 March 00. Received 2014 March 00; in original form 2014 December 00}

\newcommand{\cgs}{ ${\rm erg~cm}^{-2}~{\rm s}^{-1}$ }
\newcommand{\lum}{\rm erg~s$^{-1}$ }
\newcommand{\ha}{\rm H$\alpha$ }
\newcommand{\lo}{\rm $\lambda$ Ori }
\newcommand{\s}{\rm [S{\sc II}]/\ha }
\newcommand{\n}{\rm [NII]/\ha }
\newcommand{\ox}{\rm [OIII]/\ha }
\newcommand{\he}{\rm [HeI]/\ha }
\newcommand{\apj}{\rm ApJ}
\newcommand{\mnras}{\rm MNRAS}
\newcommand{\pasa}{\rm PASA}
\newcommand{\araa}{\rm ARAA}
\newcommand{\apjs}{\rm ApJS}
\newcommand{\apjl}{\rm ApJL}
\newcommand{\aj}{\rm AJ}
\newcommand{\pasp}{\rm PASP}
\newcommand{\aap}{\rm AAP}

\title[Scattered light in the DIG]{Models of Diffuse \ha in the Interstellar Medium: The Relative Contributions from In Situ Ionisation and Dust Scattering}
\author[Barnes et al.] {Joanna E.~Barnes$^{1}$\thanks{email: jb652@st-andrews.ac.uk}, Kenneth Wood$^1$, 
Alex S. Hill$^{2,3}$, L. Matthew Haffner$^{4,5}$
\newauthor    
\\
$^1$School of Physics \& Astronomy, University of St Andrews, North Haugh,
St Andrews, Fife, KY16 9SS, Scotland\\
$^2$CSIRO Astronomy \& Space Science, Marsfield, NSW, Australia\\
$^3$Current address: Departments of Physics and Astronomy, Haverford College, Haverford, PA USA\\
$^4$Department of Astronomy, University of Wisconsin %
 Madison, 475 North Charter Street, Madison\\
$^5$  Space Science Institute, 4750 Walnut Street, Suite 205, Boulder, CO 80301}

\maketitle

\begin{abstract}

Using three dimensional Monte Carlo radiation transfer models of photoionisation and dust scattering, we 
explore different components of the widespread 
 diffuse H$\alpha$ emission observed in the interstellar medium of the Milky Way and other galaxies. We investigate the relative contributions of H$\alpha$ from recombination emission in ionised gas and H$\alpha$ that originates in H{\sc ii} regions near the Galactic midplane and scatters off high altitude dust in the diffuse interstellar medium. For the radiation transfer simulations we consider two geometries for the interstellar medium: a three dimensional fractal geometry that reproduces the average density structure inferred for hydrogen in the Milky Way, and a density structure from a magneto hydrodynamic simulation of a supernovae driven turbulent interstellar medium.  Although some sight lines that are close to H{\sc ii} regions can be dominated by scattered light, overall we find that less than $\sim 20\%$ of the total \ha intensity in our simulations can be attributed to dust scattering. Our findings on the relative contribution of scattered H$\alpha$ are consistent with previous observational and theoretical analyses.
We also investigate the relative contributions of dust scattering and in situ ionisation of high density dust clouds in the diffuse gas.  Dust scattering in these partially ionised clouds contribute $\sim 40\%$ to the total intensity of H$\alpha$.  
 
\end{abstract}

\section{Introduction}

Widespread diffuse \ha emission is observed along all sight lines in the Milky Way and is ubiquitous in the interstellar medium (ISM) of other galaxies (see the extensive review by \citealt{Haffner2009}). The H$\alpha$ intensity is attributed to recombination emission from diffuse ionised gas (DIG, also known as the Warm Ionised Medium) in the ISM (e.g., \citealt{Miller1993}, \citealt{Dove1994}) with a contribution from H$\alpha$ photons that originate in H{\sc ii} regions close to the Galactic midplane, and are scattered towards us  by dust in the diffuse ISM (e.g., \citealt{Jura1979}, \citealt{WoodReynolds1999}). Power requirements for the DIG point to OB stars as the most likely source of its ionising photons (e.g., \citealt{Reynolds1990}). Observational and theoretical studies suggest that dust scattering is a relatively small component of the diffuse H$\alpha$ emission, typically less than 20\% of the total. 

\cite{Reynolds1988} studied the \s optical line ratio and found it to be larger in the Galactic DIG than in H{\sc ii} regions. If scattered light were the dominant source of H$\alpha$ and [S{\sc II}] in the DIG then the line ratios would be similar to those observed in H{\sc ii} regions. In addition, observations of the DIG in the Milky Way and other galaxies show that line ratios of \s and \n increase with distance from the galactic midplane (e.g., \citealt{Haffner1999}, \citealt{Rand1998}, \citealt{Otte2002}, \citealt{Hill2014}). 
 Observations of the Perseus Arm show that \s increases with decreasing \ha intensity \citep{Haffner1999}.  In analysing recent observations of the Scutum-Centaurus Arm, \cite{Hill2014} found that the increase in the \s line ratios better correlate with decreasing \ha intensity than with height above the plane, suggesting that the trend in line ratios and accompanying changes in physical conditions are primarily a function of density, not the ionising radiation field. 
The observed line ratios can be attributed to an increase in the temperature of the gas and has led to the suggestion that heating beyond that provided by photoionisation is acting in the DIG  (e.g., \citealt{Reynolds1999}, \citealt{Haffner2009}, \citealt{BlandHawthorn1997}). 

\cite{Seon2012} suggested that the elevated line ratios present in the ionised-neutral transition zone towards the edges of H{\sc ii} regions could be responsible for the observed line ratios in the DIG,  if dust scattering of light from H{\sc ii} regions was a large component of the DIG emission.  However the transition zone comprises a very small component of the total emission from the H{\sc ii} region, so if dust scattering were dominant in the DIG we would expect H{\sc ii} region-like line ratios, as originally discussed by \cite{Reynolds1988}.  

It has also been suggested that stellar \ha absorption lines are able to impact on observed emission lines, substantially increasing \s and \n line ratios in the DIG \citep{Seon2012}.
Other stellar absorption lines not coincident with ISM emission lines are also seen in WHAM (Wisconsin \ha Mapper) observations when the one-degree beam contains bright stars (V $<$ 7 mag).  However, these lines are not detectable in the $\sim 90\%$ of observations that contain only fainter sources, especially at moderate to high latitudes where diffuse stellar light is negligible.  
   Further, \cite{Otte2001} and \cite{Otte2002} considered the effects of stellar absorption lines in edge on galaxies and found no need for correction in DIG regions, only in those regions closer to the midplane.  Finally, the scale height of stellar absorption and \ha emission are not comparable, and therefore the effect of stellar absorption lines is unable to explain the observed increase with height in line ratios of \n and [SII]/H$\alpha$.  

Using observations of the high altitude cloud LDN 1780, \cite{Witt2010} derived a relation between the intensities of H$\alpha$ and $100\mu$m thermal dust emission to estimate the dust-scattered \ha contribution.  
They extrapolated their results for the LDN 1780 cloud to the high latitude ($|b| > 10^{\circ}$) sky to determine that the most probable scattered \ha intensity (0.1~R\footnote{1 R = $10^6/4\pi$ H$\alpha$ photons cm$^{-2}$ s$^{-1}$ sr$^{-1}$}) is about 19\% of the most probable total \ha intensity in this portion of the sky (0.52R).  This estimate for the dust-scattered H$\alpha$ intensity agrees with estimates from  \cite{WoodReynolds1999}, \cite{Reynolds1973} and \cite{Brandt2012}.

From a theoretical perspective, Monte Carlo scattering simulations by \cite{WoodReynolds1999} using a smooth ISM density structure (and assumed \ha emissivity from the DIG) found that less than $20\%$ of the total \ha intensity from the DIG is a result of dust-scattered \ha from H{\sc ii} regions. Their simulations showed spatial variations, with the scattered light component being smallest at high altitudes and with some sight lines towards the galactic midplane exhibiting a much larger scattered light component. The results from the smooth density ISM models of \cite{WoodReynolds1999} are in broad agreement with other estimates of the dust-scattered contribution to the observed diffuse H$\alpha$ (e.g., \citealt{Reynolds1988}).   If scattered light were a significant contributor to the observed \ha intensity it would complicate the interpretation of the observed \ha intensity as a tracer of the electron density along the line of sight.   This would have significant implications for the understanding of both the energy transport in the ISM and for the use of \ha as a template for the Galactic foreground free-free contribution to the cosmic microwave background.  

In this paper we extend the work of \cite{WoodReynolds1999} to study H$\alpha$ emission and scattering in three dimensional ISM density structures. Our models employ three dimensional (3D) Monte Carlo radiation transfer codes to compute the photoionisation and temperature structure of the DIG and thus the 3D H$\alpha$ emissivity from in situ recombinations.  We then use a separate scattering code to compute the total intensity of H$\alpha$ from in situ recombinations and H$\alpha$ that originates in H{\sc ii} regions and is scattered by dust in the diffuse ISM. The setup of our simulations and methods are outlined in section \ref{models}, our results are presented in section \ref{results}, and our conclusions are presented in section \ref{conclusions}.

\section{Models}
\label{models}
\subsection{Photoionisation Models}
For our study of photoionisation and scattering in the DIG we adopt two density structures. First we consider a subsection of a supernova driven, magnetohydrodynamic (MHD) simulation of the ISM that extends to $|z| = \pm 2$~kpc with width 1~kpc \citep{Hill2012}.  The density in these simulations is strongly peaked around the midplane and has a small scale height, such that the density above $\sim 300$pc is smaller than inferred in the Galaxy. The MHD simulations include type Ia and core collapse supernovae set off at the average galactic supernova rate without knowledge of the gas distribution.  They do not include photoionisation, therefore we post process the density grids using our photoionisation and scattered light codes. 

Due to the small density scale height in the MHD simulations, we also consider a fractal density structure that has a vertical density distribution closer to that inferred for our Galaxy \citep{Barnes2014}.  To create this model ISM we convert a smooth four-component density distribution to a fractal structure (see below). The smooth density 
comprises a Dickey-Lockman distribution \citep{DL1990} plus an extended component:
\begin{eqnarray}
\nonumber n(z) = 0.4e^{-(|z|/90 )^{2}/2}+ 0.11e^{-(|z|/225 )^{2}/2}\\
+ 0.06e^{-|z|/400 }+0.04e^{-|z|/1000}
\label{eq1}
\end{eqnarray}
where the height $z$ is in pc and number densities are in cm$^{-3}$.  

When converted to a 3D fractal structure, this density is the input for our Monte Carlo photoionisation and scattering simulations.  The gas is initially assumed to be neutral and at the end of the photoionisation simulations 
will comprise ionised and neutral components.  The first three terms in equation \ref{eq1} represent a Dickey-Lockman distribution for the average density of the neutral hydrogen.  The fourth component is more vertically extended, and for fractal models is almost fully ionised at the end of our photoionisation simulation, thus representing the density of the warm ionised medium with a 1 kpc scale height \citep{Haffner1999}.
 
To allow ionising photons to propagate and reach gas at high altitudes, we convert the smooth structure to a fractal one using the algorithm of \cite{Elmegreen1997} as described in \cite{wood2005} where this algorithm has been scaled to the box size used here (1kpc $\times$ 1kpc $\times$ 4kpc).  We adopt a five-level clumping algorithm and arrange the density structure such that one third of the mass is distributed smoothly, with the remainder in fractal clumps.  The fractal algorithm maintains the total mass and average density with height.  In both the smooth and fractal photoionisation simulations, we begin by assuming all of the hydrogen is neutral, and then allow the gas to be ionised.

The resolution in both the MHD and fractal ISM models is 15.6pc per grid cell and we therefore do not resolve traditional parsec-scale H{\sc ii} regions around OB stars.  We have investigated photoionisation simulations with higher resolution and find there is little difference in the large-scale ionisation and temperature structure between high and low resolution runs.

To determine the relative contribution of \ha from in situ recombination of ionised hydrogen in the DIG, we compute the 3D \ha emissivity using a Monte Carlo photoionisation code \citep{Wood2004}.  Due to the grid resolution described above, the sources of ionising radiation in our simulations represent photons escaping from H{\sc ii} regions.  We reproduce the Galactic surface density of O stars in the Solar neighbourhood \citep{Garmany1982} by randomly placing 24 sources in the $xy$ plane with a scale height of 63 pc in $|z|$ \citep{Maiz2001}. Since our simulations do not resolve the H{\sc ii} regions, we treat such regions as ``point sources'' located at the source positions.  In what follows these sources will be referred to as ``H{\sc II} regions''.  

The spectrum of the sources in our ionisation simulations is assumed to be that of a typical O star with $T = 35000 \mathrm{K}$.  Although altering the type of O star will lead to small changes in the temperature of the gas (and therefore the \ha emissivity) the most important variable is the ionising luminosity \citep{Wood2010}.

We find from our ionisation simulations that the ionising luminosity that produces the extended DIG is $Q = 1 \times 10^{49}$ s$^{-1}$ for the MHD and $Q = 1.6 \times 10^{50}$ s$^{-1}$ for the fractal structure. 
Our photoionisation code computes the gas temperature and ionisation state of H, He, C, S, N, O and Ne in each cell.   For an input density structure we calculate the ionisation and temperature structures arising from photoionisation only, without considering photoelectric or shock heating, the two major heating mechanisms in the MHD simulations.  The 3D H$\alpha$ emissivity of the DIG then follows from our photoionised density grid. 

Observations of ionised gas in other galaxies indicate that the \ha emission from the DIG is approximately equal to that from traditional  H{\sc ii} regions (e.g., \citealt{Ferguson1996}, \citealt{Zurita2000}, \citealt{Thilker2002}, \citealt{Oey2007}).  We therefore calculate the \ha luminosity of the DIG in the photoionisation simulations and set the \ha luminosity from the point sources equal to this. For the MHD simulations the total H$\alpha$ luminosity from H{\sc ii} regions, is $L_{{\rm H}\alpha} = 2.5\times 10^{48}\, {\rm s}^{-1}$ and for the fractal simulations it is $L_{{\rm H}\alpha} = 6.5\times 10^{49}\, {\rm s}^{-1}$. We adopt this approach for assigning the H$\alpha$ luminosity from H{\sc ii} regions, but note that a fully self-consistent model would require sub-grid resolution to compute the ionisation structure and resulting H$\alpha$ luminosity from the H{\sc ii} regions as well as the diffuse ionised gas. 

 The H$\alpha$ emission in the DIG is a result of recombinations in the ionised gas.  The number of Lyman continuum photons that can reach the DIG and ionise it depends on the number of photons that are able to escape H{\sc ii} regions and how many of these photons then escape the galaxy. 

 It is thought that globally 5\% of the Lyman continuum photons from OB stars escape the Galaxy (e.g. \citealt{Kim2013}, \citealt{Barger2013}), 15\% produce the DIG  (\citealt{Reynolds1990}) and the remaining 80\% produce local H{\sc ii} regions close to sources. However there is certainly local variation in these fractions, which we explore by varying the luminosity available to ionise the DIG versus H{\sc ii} regions. 
 \cite{Rogers2013} used hydrodynamic models of massive star clusters to estimate that the percentage of ionising photons that escape the cluster increases with age from 1\% to 60\% over the first 4Myr of the cluster's evolution.
 Therefore, in addition to a model where the \ha flux from the DIG and H{\sc ii} regions are equal, we also investigate
  simulations where 30\% of the \ha flux originates in H{\sc ii} regions and 70\% in the DIG, and simulations where 70\% of the \ha flux originates in H{\sc ii} regions and 30\% in the DIG.

\subsection{Scattered Light Models}
\label{models:scatter}

To simulate the scattering of \ha photons we use the Monte Carlo scattering code described by \cite{WoodReynolds1999}. We assume that the dust and gas are well mixed and represented by a \cite{Mathis1977} mixture with total opacity $\kappa = 220$ cm$^2$g$^{-1}$ and scattering albedo $a= 0.5$ appropriate for H$\alpha$ photons.  To describe the angular shape of the dust scattering we use a forward throwing Henyey-Greenstein phase function HG($\theta$) with anisotropy parameter $g=0.44$,

\begin{equation}
HG(\theta) = \frac{1}{4\pi}\frac{1-g^{2}} {(1+g^{2}-2g\cos\theta)^{3/2}}
\end{equation}
We use a ``forced first scattering''  so every photon contributes to the scattered light intensity and a ``peeling off'' algorithm forcing photons towards the observer with appropriate weights.  We adopt the forced first scattering procedure to investigate scattered light in optically thin gas.  Our scattered light models simulate 
the scattering of H$\alpha$ photons that originate in both H{\sc ii} regions and the DIG.
\section{Results}
\label{results}

The Monte Carlo scattering code computes the H$\alpha$ intensity comprising photons that reach the observer without scattering from the DIG and point source H{\sc ii} regions. The code also computes the contributions from H$\alpha$ photons that originate in the DIG and H{\sc ii} regions and are scattered into our line of sight.  Therefore there are four types of H$\alpha$ photons present in our simulations:
\begin{enumerate}
\item {Those that originate in H{\sc ii} regions and reach the observer without scattering}
\item {Those that originate in H{\sc ii} regions and scatter off dust in the DIG before reaching the observer}
\item {Those that originate in the DIG from recombinations and reach the observer without scattering}
\item {Those that originate in the DIG from recombinations and scatter off dust before reaching the observer}
\end{enumerate}

Because we wish to determine the relative contribution to the \ha intensity from photons that originate in H{\sc ii} regions and scatter in the DIG, we hereafter refer to case (ii) photons as ``scattered light'' and photons in cases (iii) and (iv) as``diffuse''.  We do not differentiate between H$\alpha$ photons that originate in the DIG and reach the observer without scattering and those that originate in the DIG and are subsequently scattered.

\subsection{Edge-on viewing}
\begin{figure*}
\begin{center}
\includegraphics[scale=1,trim=0mm 120mm 30mm 20 mm,clip]{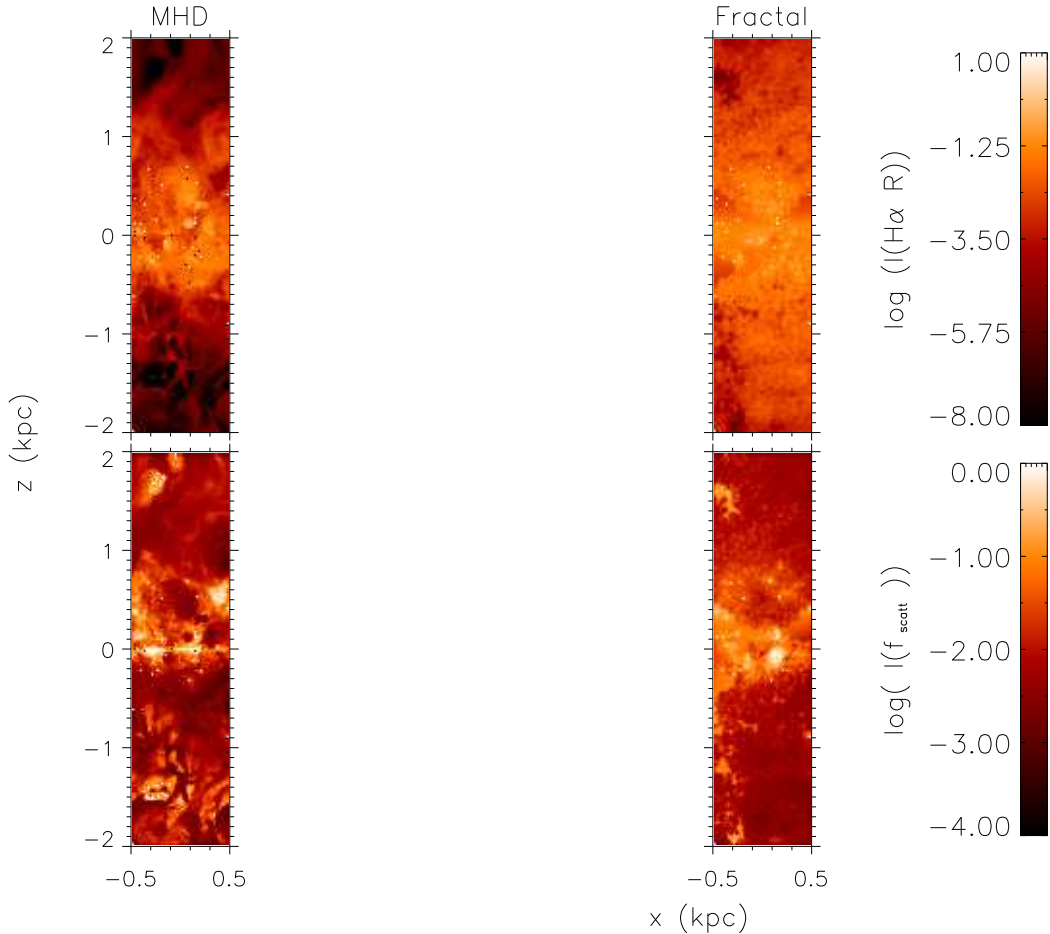}
\caption{Maps showing total \ha intensity (top row) and fraction of scattered light that originates in HII regions (bottom row).   Left: MHD simulations, right: fractal models using HG scattering phase function. }
\label{maps}
\end{center}
\end{figure*}

The upper panels of figure \ref{maps} show the total \ha intensity (photons that originate in the DIG plus those from H{\sc ii} regions including scattered photons) and the lower panels show the ratio of scattered to total intensity (photons from case (ii)/total \ha intensity).  The fraction of scattered light appears to decrease with distance from the midplane, however there are regions where the contribution from scattered light is large, particularly noticeable in the MHD simulations. 

 The fractal density structure (right panels of figure \ref{maps}) has a larger density scale height than that of the MHD simulations.  This density structure is less centrally peaked than the MHD simulations and has higher density at large $|z|$ (see figure 1 in \citealt{Barnes2014}).  The right panel of figure \ref{maps} shows maps of the total \ha intensity and the ratio of scattered to total intensity in a fractal density structure. 
 Comparing the fractal models to the MHD density grid, we see that the total \ha intensity is higher above the midplane in the fractal models and the overall fraction of intensity that is dust scattered from H{\sc ii} regions is smaller.  

The scattered light contribution to the total \ha emissivity scales as column density ($\sim n$) while the contribution from photoionisation scales as $n^2$. At low densities dust scattering may therefore contribute a large fraction of the total H$\alpha$ intensity. This results in the smaller fraction of scattered light in the fractal models.

\begin{figure*}
\begin{center}
\includegraphics[scale=1,trim=15mm 100mm 0mm 250 mm,clip]{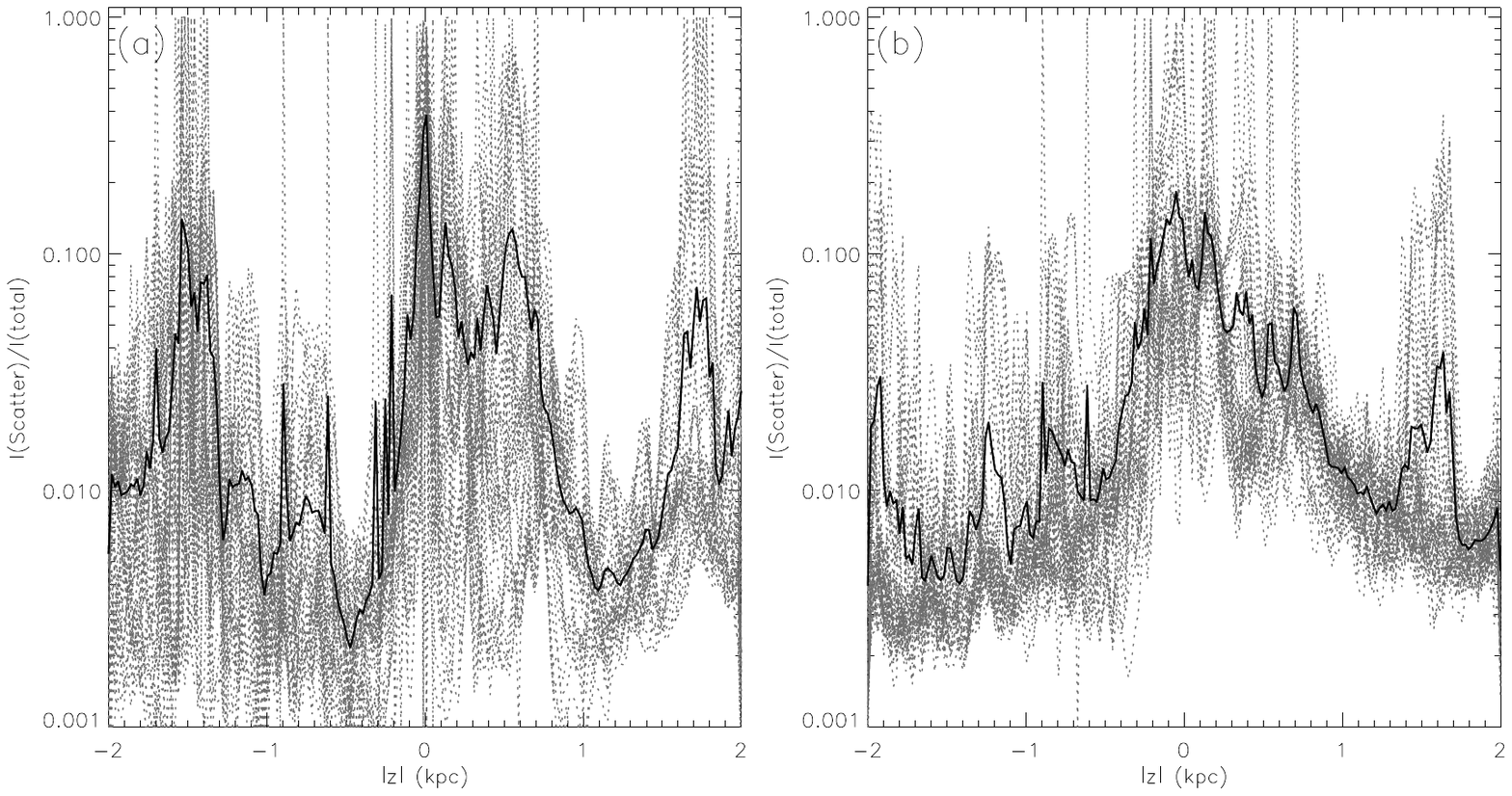}
\caption{Ratio of scattered to total light vs distance from midplane.  The solid line shows the average ratio and the dotted grey lines show every sight line through the simulations box for (a) MHD simulations, (b) Fractal simulations with HGG phase function. }
\label{lines}
\end{center}
\end{figure*}

 
Figure \ref{lines} shows the fraction of scattered light intensity to total intensity for every 1 pixel wide slice (grey lines) through the simulation box and the average fraction (black line). 
  We find that on average the largest fraction ($\sim 40\%$) of scattered light is located close to the midplane of the simulation, where the density is highest.  This is a result of the close proximity of the midplane dust to the H{\sc ii} regions and the $1/r^2$ dependence of scattered light.  In the MHD simulations the fraction of scattered light then decreases as the density decreases to below $\sim 10\%$ above $|z| \sim 300$pc and below $\sim 5\%$ in the fractal models.  

The large peaks in scattered light fraction above the midplane in the MHD simulations arise because of the very low density in individual cells resulting in low DIG emissivity.  Since the DIG emissivity scales as $\sim n^{2}$ and scattered light scales as $n$, scattered light will dominate in these low density cells.  However in the MHD simulations the total \ha intensity in the majority of the low density regions is extremely faint and below the WHAM detection limit of 0.1R.
 
\begin{figure*}
\begin{center}
\includegraphics[scale=0.75,trim=18mm 40mm 0mm 50 mm,clip]{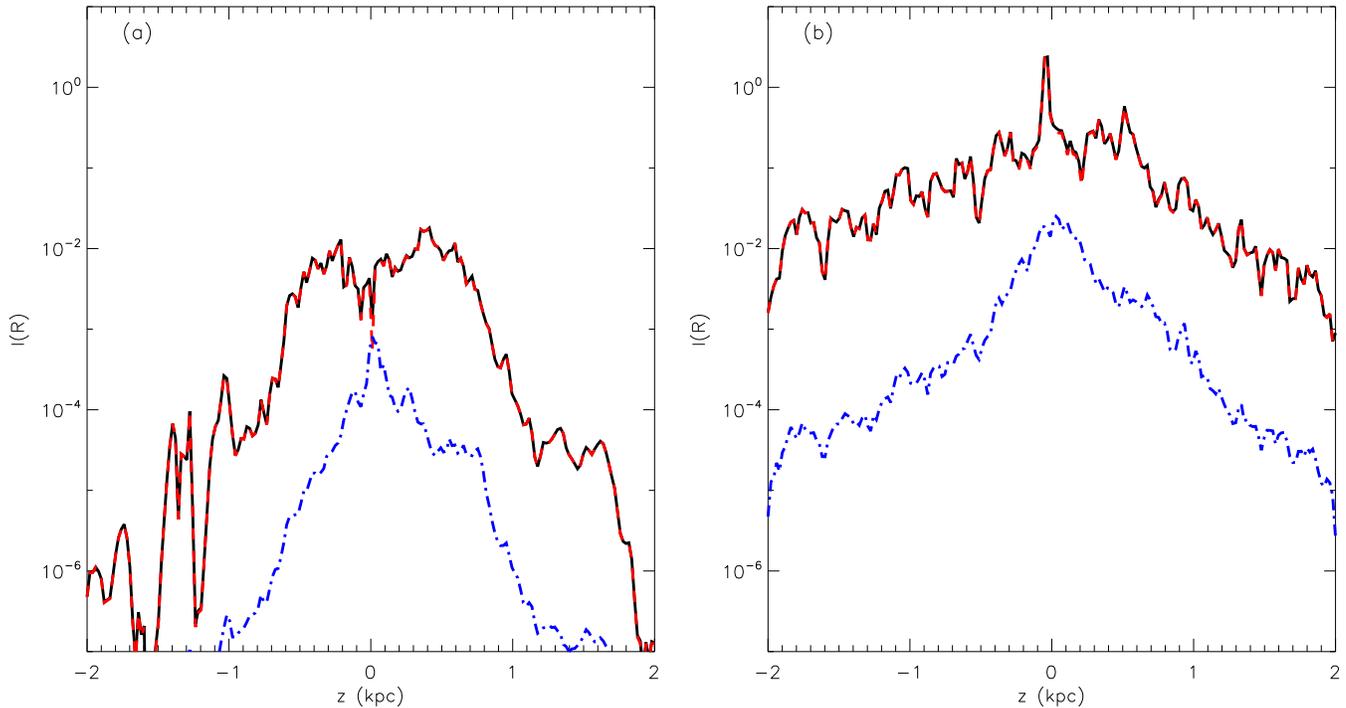}
\caption{Intensity of total \ha (black), \ha in a 1 pixel wide slice from photoionised DIG (red dashed), intensity of \ha from HII regions scattered in to the DIG (blue dot-dashed) for (a) MHD simulations, (b) fractal simulations with HG phase function. }
\label{comparison}
\end{center}
\end{figure*}

Figure \ref{comparison} shows the total \ha intensity (black), intensity of \ha originating in the DIG (red) and intensity of dust-scattered \ha from H{\sc ii} regions (blue) for a one pixel wide slice through the MHD and fractal simulations.  The intensities of both the DIG and scattered light from HII regions peak close to the midplane of the simulation, where the density is highest and the dust is closest to the sources.  


We have investigated models where we distribute the \ha luminosity such that 30\% of the total \ha flux originates in H{\sc ii} regions.  The peak fraction of scattered light in these simulations decreases by between $\sim 10\%$ and 20\%. If we distribute the \ha flux so 70\% originates in HII regions then the peak fraction of scattered light increases by 10\%.  In both cases the fraction of scattered light still decreases to below $\sim 20\%$ in the MHD simulations.  However when the majority of the \ha flux originates in H{\sc ii} regions, the fraction of scattered light above the midplane increases to $\sim 10\%$ in the fractal models.   

We have also investigated simulations using different albedo and scattering phase functions ($a=0.67,0.77$ and $g = 0.5,0.55$ taken from \citet{Weingartner2001}).  We find that altering the albedo and scattering phase function has qualitatively little effect on our simulations, with the greatest difference occurring when $a=0.77$ and $g = 0.5$ where the fraction of scattered light is increased to $\sim 15\%$ above the midplane in MHD simulations.  However this does not significantly alter our results with the largest average fraction of scattered light increasing from $40\%$ to $50\%$. 

\subsection{Face-on viewing}

\begin{figure*}
\begin{center}
\includegraphics[scale=1,trim=18mm 120mm 60mm 50 mm,clip]{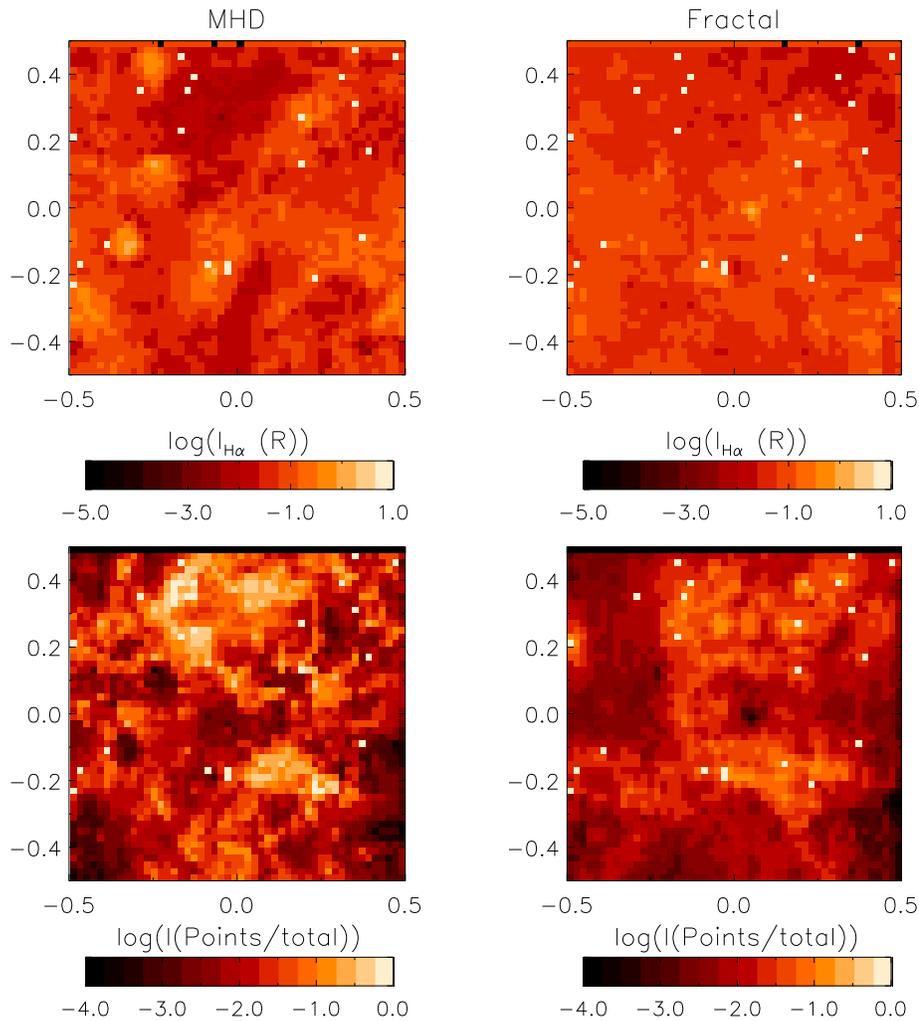}
\caption{Maps showing total \ha intensity (top row) and fraction of scattered light that originates in HII regions (bottom row) for face on views of our simulations. Left: MHD simulations and right: fractal simulations. }
\label{faceonmaps}
\end{center}
\end{figure*}

\begin{figure*}
\begin{center}
\includegraphics[scale=0.75,trim=18mm 40mm 0mm 50 mm,clip]{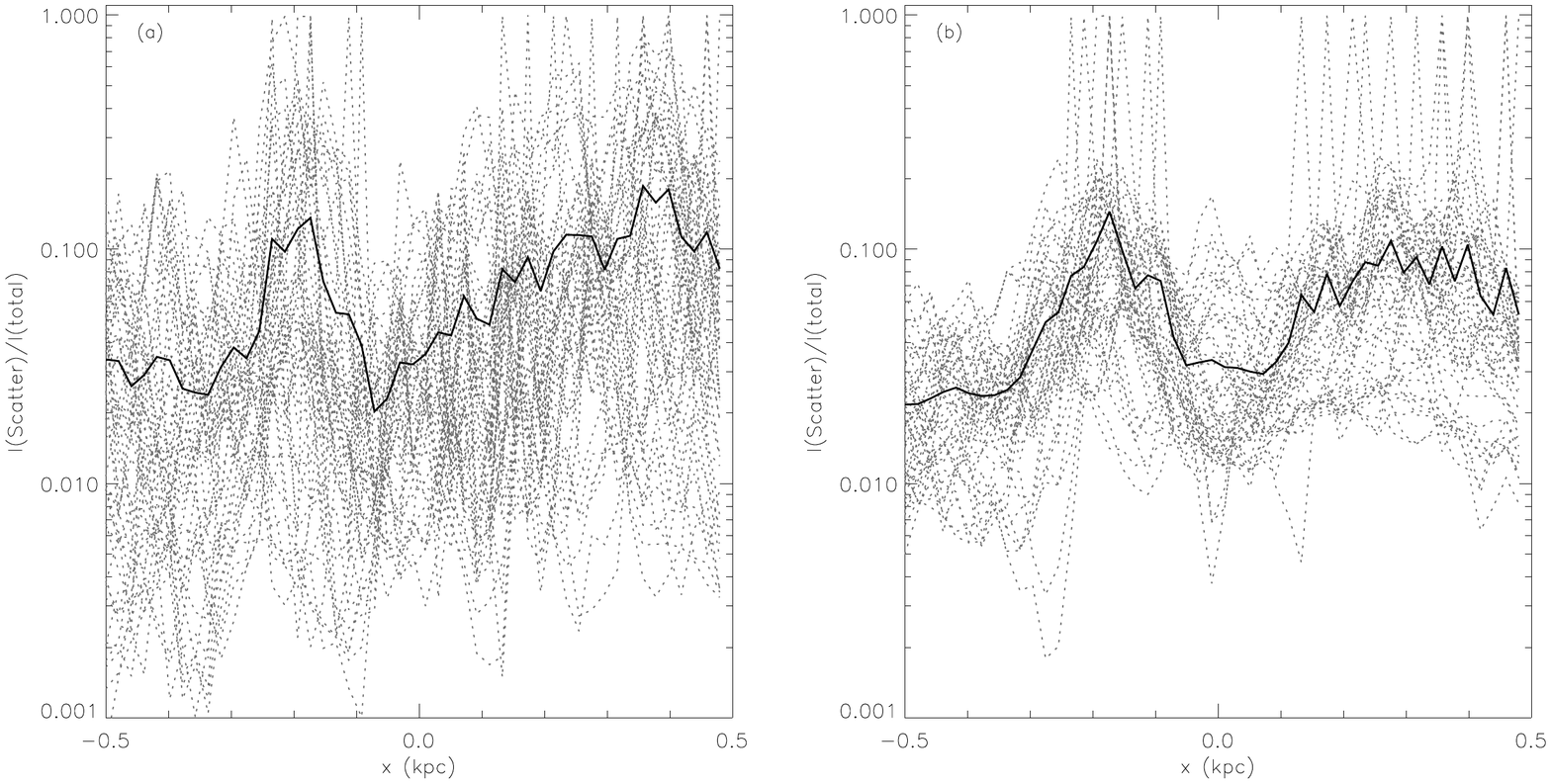}
\caption{Face on view of the ratio of scattered to total light vs distance from midplane.  The solid line shows the average ratio and the dotted grey lines show every sight line through the simulations box for (a) MHD simulations and (b) Fractal simulations with HG phase function}
\label{faceonlines}
\end{center}
\end{figure*}

Figures \ref{faceonmaps} and \ref{faceonlines} show a face on view of the \ha intensity of scattered light and the ratio of scattered light to total intensity. 
Face-on and edge-on viewing gives ratios of scattered to total H$\alpha$. 
Figure \ref{faceonlines} shows that in the MHD simulations the fraction of scattered light is typically below 20\% while it is below 10\% in the fractal models, consistent with our results from edge on visualisations and with previous results (\citealt{Reynolds1988}, \citealt{WoodReynolds1999}, \citealt{Ferrara1996})

\subsection{Galactic Cirrus}

High-density clouds hundreds of parsecs above the midplane could have an effect on the fraction of total \ha intensity due to scattered light.  Observations of the cloud LDN1780 led \citet{Witt2010} to determine that dust-scattered \ha accounts for $\sim19\%$ of the \ha emission at high altitudes in the Galaxy.  LDN 1780 is located approximate 110 pc above the midplane of the galaxy \citep{Franco1989}, is approximately 1.2 pc in diameter and has an average density of $\sim 10^3$~cm$^{-3}$. To determine the contribution of dust-scattered light from such clouds we estimate the thickness and \ha intensity of the ionised shell that results from photoionistion using the following analysis.  We assume a slab geometry for the cloud and plane parallel illumination.

The ionising luminosity needed to ionise a volume $\delta V$ is determined by 
\begin{equation}
Q = n^{2}\alpha_{B}\delta V
\end{equation}
where Q is the number of ionising photons per second incident on volume $\delta V$, $n$ is the density of the gas (cm$^{-3}$), $\alpha_{B}$ is the recombination coefficient assuming Case B recombination.  This can also be written in terms of the total ionising flux available from O stars in the galaxy $F_{\rm LyC}$:
\begin{equation}
F_{\rm LyC}f\delta A = n^{2}\alpha_{B}\delta A\delta l
\end{equation}
where $\delta A$ is the area of the cloud exposed to the ionising radiation, $f$ is the fraction of ionising photons that escape H{\sc ii} regions to produce the DIG, and $\delta l$ is the depth of the ionised volume in the cloud.  Given an ionising flux, the depth to which a cloud can be ionised is:
\begin{equation}
\delta l = \frac{F_{\rm LyC}f}{n^{2}\alpha_{B}}
\end{equation}
The intensity of \ha emission in Rayleighs is related to the emission measure by 

\begin{equation}
EM = \int {n^{2}dl} = 2.75 T_{4}^{0.9}I_{H\alpha}(R) \mbox{ cm}^{-6}\mbox{pc}
\end{equation}
\citep{Haffner1999}.  The \ha intensity can therefore be found using
\begin{equation}
 I_{H\alpha}(R) = \frac{\int{n^{2}dl}}{2.75T _{4}^{0.9}}
\end{equation}
assuming the density is constant along the path $dl$ this can be simplified to 
\begin{equation}
I_{H\alpha}(R) = \frac{n^{2}\delta l}{2.75T _{4}^{0.9}}
\end{equation}
where $T_{4}$ is the temperature of the gas in units of $T/10^4$K and $I_{H\alpha}$ is the intensity of \ha in Rayleighs (R).  
 
Adopting $f=0.15$, we would expect a cloud with  
density $n = 10^3$cm$^{-3}$ and $T = 10^4$ K, ionised by the galactic ionising flux from O stars, $F_{\rm LyC} = 3 \times 10^{7}$cm$^{-2}$s$^{-1}$\citep{Reynolds1995},  would be ionised to a depth of $6 \times 10^{-6}$pc and produce an H$\alpha$ intensity of $ 2.2$R from in situ recombinations.

Since we are unable to resolve an ionised skin of this thickness in the large-scale simulation presented in section 2, we explore scattering on smaller scales with a model of a single cloud.  The total Lyman continuum flux in the Galaxy is $3.74 \times 10^{7}$ cm$^{-2}$s$^{-1}$ \citep{Vacca1996}.  We assume that half of this flux travels upwards from the midplane of the Galaxy, towards the cirrus cloud, while the other half travels downwards, away from the cloud.  We assume that $5\%$ of the Lyman continuum photons from each source escape the galaxy (e.g. \citealt{Kim2013}, \citealt{Barger2013}), $15\%$ produce the DIG \citep{Reynolds1990} and the remaining 80\% produce local H{\sc ii} regions around each source. 
We are considering the scattering of \ha photons that originate in H{\sc ii} regions, which are produced by 80\% of the total ionising luminosity.   Assuming case B recombination, each Lyman continuum photon produces 0.46 \ha photons \citep{Martin1988}.  Therefore the \ha flux impinging on galactic cirrus clouds is  $0.46 \times 0.8F_{LyC} = 0.37F_{LyC}=6.9 \times 10^{6}$cm$^{-2}$s$^{-1}$. 


 To determine the contribution of dust scattering in this cloud, we create simulations of a spherical cloud with $r =0.5$pc and $n = 10^3$cm$^{-3}$ using a $200^3$ pixel grid.  The H$\alpha$ flux incident on the cloud in this simulation is assumed to be directed upwards from H{\sc ii} regions close to the midplane of the Galaxy.  We then run the scattered light simulations described above and find that the intensity of H$\alpha$ scattered by the cloud is 1.4R, which is about 40\% of the total H$\alpha$ intensity from the cloud.
These results indicate that the presence of high density galactic cirrus can increase the contribution of dust scattered light to the total \ha intensity we observe in the Galaxy.  However the intensity of \ha emission that results from ionisation is still larger than that from scattering.

\section{Conclusions}
\label{conclusions}

Using MHD and analytic fractal models for the 3D density structure appropriate for the ISM in the outer disk of a spiral galaxy, we have investigated the relative contributions to the H$\alpha$ intensity from in situ recombinations of diffuse ionised gas and dust-scattered H$\alpha$ originating in H{\sc ii} regions. Our models self consistently compute the diffuse H$\alpha$ emissivity from diffuse ionised gas. We do not resolve small scale H{\sc ii} regions within our photoionisation simulations, so make the assumption that the H$\alpha$ luminosity from H{\sc ii} regions is equal to what we compute from the DIG. The main results of our combined photoionisation and H$\alpha$ scattered light models are:

\begin{itemize}

\item   The intensity of scattered \ha originating from H{\sc ii} regions differs depending on the density structure.  In both fractal and MHD structures the intensity of scattered light peaks around the midplane of the simulation closest to the H{\sc ii} regions and where the gas density is highest.  The intensity of scattered light then decreases away from the midplane to less than about $\sim 10\%$ in the MHD and $\sim 5\%$ in fractal models. The larger scattered light fraction in the MHD simulations is due to the very low densities and hence low intrinsic H$\alpha$ emissivity at large heights in those models.

\item In low density regions a large fraction of the \ha in our simulations is dust scattered light that originates in H{\sc ii} regions, a result of the small \ha emissivity from the lowest density DIG.  

\item Different scattering phase functions and albedo affect the intensity of scattered light in the models, however this does not significantly change our results, increasing the largest fraction of scattered light  by 10\%.

\item  Scattering of \ha photons from H{\sc ii} regions off high density cirrus in the ISM can dominate over the \ha intensity from photoionisation, contributing 40\% of the total H$\alpha$.  However the covering fraction of such clouds is $\sim 50\%$ \citep{Gillmon2006}, so such clouds would not effect all sight lines through the Galaxy.

\end{itemize}

\section*{acknowledgments}
The authors would like to thank Kwang Il-Seon and Adolf Witt for their helpful comments on an early version manuscript.  JB acknowledges the support of an STFC studentship.  LMH acknowledges support from the U.S. National Science Foundation through award AST-1108911.  
\bibliographystyle{mn2e}
\bibliography{scatterarchive}
\end{document}